\documentclass[]{spie}
\usepackage{aas_macros}
\def\spider{{\sc Spider}}
\def\boom{{\sc Boomerang}}
\def\blast{{\sc BLAST}}
\def\bicep{{\em Robinson/BICEP}}

\newcommand{\planck}{{\it Planck}}

\def\mukcmbrts{\mu\mathrm{K}_{\mathrm{\mbox{\tiny\sc cmb}}}\sqrt{\mathrm{s}}}

\newcommand{\sini}{$\mathrm{Si}_3\mathrm{N}_4$ }

\usepackage[]{graphicx}

\title{SPIDER: a Balloon-borne Large-scale CMB Polarimeter}

\author{B. P. Crill\supit{a,b}, P.A.R. Ade\supit{c},
E. S. Battistelli\supit{d}, S. Benton\supit{e}, R. Bihary\supit{f}, 
J. J. Bock\supit{g,h}, J. R. Bond\supit{b}, J. Brevik\supit{h}, S. Bryan\supit{f}, 
C. R. Contaldi\supit{i}, O. Dor{\'e}\supit{b}, M. Farhang\supit{a}, 
L. Fissel\supit{a}, S. R. Golwala\supit{h}, 
M. Halpern\supit{c}, G. Hilton\supit{j}, W. Holmes\supit{g}, V. V. Hristov\supit{h}, 
K. Irwin\supit{j}, W. C. Jones\supit{g,h}, C. L. Kuo\supit{k}, A. E. Lange\supit{h}, 
C. Lawrie\supit{f}, C. J. MacTavish\supit{b}, T. G. Martin\supit{l}, 
P. Mason\supit{h}, T. E. Montroy\supit{f}, C. B. Netterfield\supit{a,e}, 
E. Pascale\supit{c}, D. Riley\supit{f}, J. E. Ruhl\supit{f}, M. C. Runyan\supit{h}, 
A. Trangsrud\supit{h},C. Tucker\supit{c}, A. Turner\supit{g}, 
M. Viero\supit{a}, and D. Wiebe\supit{e}.
\skiplinehalf
\supit{a}Department of Astronomy and Astrophysics, University of Toronto, Toronto, ON, Canada; \\
\supit{b}Canadian Institute for Theoretical Astrophysics (CITA),University of Toronto, ON, Canada; \\
\supit{c}School of Physics and Astronomy, Cardiff University , UK; \\
\supit{d}Department of Physics and Astronomy, University of British Columbia, Vancouver, BC, Canada;\\
\supit{e}Department of Physics, University of Toronto, ON, Canada;\\
\supit{f}Department of Physics, Case Western Reserve University, Cleveland, OH, USA;\\
\supit{g}Jet Propulsion Laboratory, Pasadena, CA, USA;\\
\supit{h}Department of Physics, California Institute of Technology, Pasadena, CA, USA;\\
\supit{i}Theoretical Physics, Blackett Laboratory, Imperial College, London, UK;\\
\supit{j}National Institute of Standards and Technology, Boulder, CO, USA;\\
\supit{k}Department of Physics, Stanford University, Palo Alto, CA, USA;\\
\supit{l}Department of Mechanical and Industrial Engineering, University of Toronto, ON, Canada
}

\authorinfo{Further author information: (Send correspondence to B.P.Crill) E-mail: {\tt crill@astro.utoronto.ca}, Telephone: $+$1 416 978 0307}

\begin{document} 
  \maketitle 

%%%%%%%%%%%%%%%%%%%%%%%%%%%%%%%%%%%%%%%%%%%%%%%%%%%%%%%%%%%%% 
\begin{abstract}
\spider\ is a balloon-borne experiment that will measure the polarization of the Cosmic Microwave Background over a large fraction of a sky at $\sim1^\circ$ resolution. Six monochromatic refracting millimeter-wave telescopes with large arrays of antenna-coupled transition-edge superconducting bolometers will provide system sensitivities of 4.2 and 3.1 $\mukcmbrts$ at 100 and 150 GHz, respectively.  A rotating half-wave plate will modulate the polarization sensitivity of each telescope, controlling systematics.  Bolometer arrays operating at 225 GHz and 275 GHz will allow removal of polarized galactic foregrounds. In a 2-6 day first flight from Alice Springs, Australia in 2010, \spider\ will map 50\% of the sky to a depth necessary to improve our knowledge of the reionization optical depth by a large factor.  
\end{abstract}

\section{INTRODUCTION}
\label{sec:intro}  

Rapid progress in millimeter-wave receiver technology enabled major advances in cosmology over the past decade.   Extremely sensitive receivers measured the very faint anisotropies and polarization of the Cosmic Microwave Background (CMB), which are a mere part in $10^5$ and $10^6$ of the background, respectively.  CMB observations from the ground, stratospheric balloons, and space have confirmed and powerfully constrained the standard model of cosmology.  Two periods of the history of the universe still elude complete description: the period of inflation in the very early universe, and the period of first star formation that reionized the intergalactic medium in the more recent universe.

The theory of inflation postulates a period of very rapid expansion of the Universe ($\sim 10^{-36}$ seconds after the Big Bang) that seeded the formation of structure.  The theory remains incomplete and current observations only provide general constraints\cite{2008arXiv0803.0547K}.  Inflation also  predicts that the universe is filled with gravitational waves generated in the early universe.  The gravitational waves have extremely long wavelength, making them difficult to detect directly, but they leave a unique imprint in the polarization pattern in the CMB: the B-mode polarization.  Measuring the B-mode signal (a part in $10^7$ of the background) requires another leap forward in technology.

The CMB also provides a way to investigate the period of reionization of the universe.  Reionization enhances polarization and anisotropies on scales larger than the horizon size at reionization and damps them on smaller scales.  Measuring the scale of damping and the amount of tells us the optical depth of reionized universe through which the CMB photons have travelled.  Finding the signature of reionization is easiest in  polarization, as other cosmological parameters can create a similar signature in the temperature anisotropy.  The 5 year results from WMAP included a measurement of the total integrated optical depth, but tighter constraints could be found with a more sensitive instrument.

Here we describe \spider, a balloon-borne instrument designed to measure the polarization of the CMB over a large fraction of the sky at large angular scales.  The optical design of \spider, including a rotating  half-wave plate at the telescope aperture and on-axis monochromatic refracting telescopes, will provide the required control of systematics.  Large arrays of antenna-coupled transition-edge superconducting bolometers with multiplexed SQUID readouts provide the raw sensitivity for probing inflation and reionization.  The choice of frequency bands in \spider\ will minimize confusion from galactic foregrounds.  \spider\ complements \planck\  with its greater control of polarization systematics and closely spaced passbands for polarized foreground  characterization.

The instrument will make a 2-6 day first flight from Alice Springs, Australia in the austral autumn of 2010 that will probe the reionization of the universe with an unprecedented measurement of the large scale polarization of the CMB.  Subsequent flights will provide the longer integration time necessary for investigating inflation.  The balloon-borne platform is an ideal environment for these observations, allowing observation of nearly half the sky.  

The \spider\ instrument was introduced in Montroy et al. (2006) \cite{2006SPIE.6267E..24M}.    MacTavish et al. (2007) \cite{2007arXiv0710.0375M} made a detailed investigation of systematics in \spider\ observations.  Here we discuss modifications in the instrument design and discuss the prospects for scientific results from our first flight. 

\section{SCIENCE GOALS}
\label{sec:science}

Polarization of the CMB comes from Thomson scattering of anisotropic radiation by electrons in the primordial plasma at the surface of last scattering, when the CMB was emitted.  No circular polarization is created by this process.  The linear polarization pattern on the full sky can be decomposed into rotationally invariant E-modes and B-modes.  These quantities are linearly related to the Stokes parameters Q and U\footnote{Q and U completely describe the polarization of the CMB, but are coordinate-dependent quantities.}.  The majority of the polarized signal in the CMB is due to the velocity of the primordial plasma and appears as a series of acoustic peaks in the E-mode angular power spectrum.  Cosmological B-mode polarization is uniquely  produced by very long wavelength gravitational waves generated the early universe (see Section  \ref{sec:inflation} below). 

\spider\rq s  first flight will be a turnaround flight\footnote{\emph{Turnaround} refers to a balloon launched at the time of the seasonal reversal of the stratospheric wind direction; the balloon drifts slower in longitude.} from Alice Springs, Australia, of 2-6 days duration in the austral autumn of 2010.    While the sun is below the horizon, \spider\ will operate in its spinning mode, and will map $\sim 50\%$ of the sky each night.  When the sun is between the horizon and $45^\circ$ elevation, \spider\ will scan a reduced section of the sky in the anti-sun direction.  During this initial flight, \spider\ will make measurements of the large scale E-mode angular power spectrum of the CMB and dust polarization at large angular scales.

Future flights of \spider\ will be Long Duration Balloon flights of 20-30 days from Antarctica or from Alice Springs and will provide the long integration time necessary to target the the signature of primordial gravitational waves in the B-mode power spectrum.

\begin{figure}[!t]
\begin{center}
\includegraphics[width=4.5in,angle=-90]{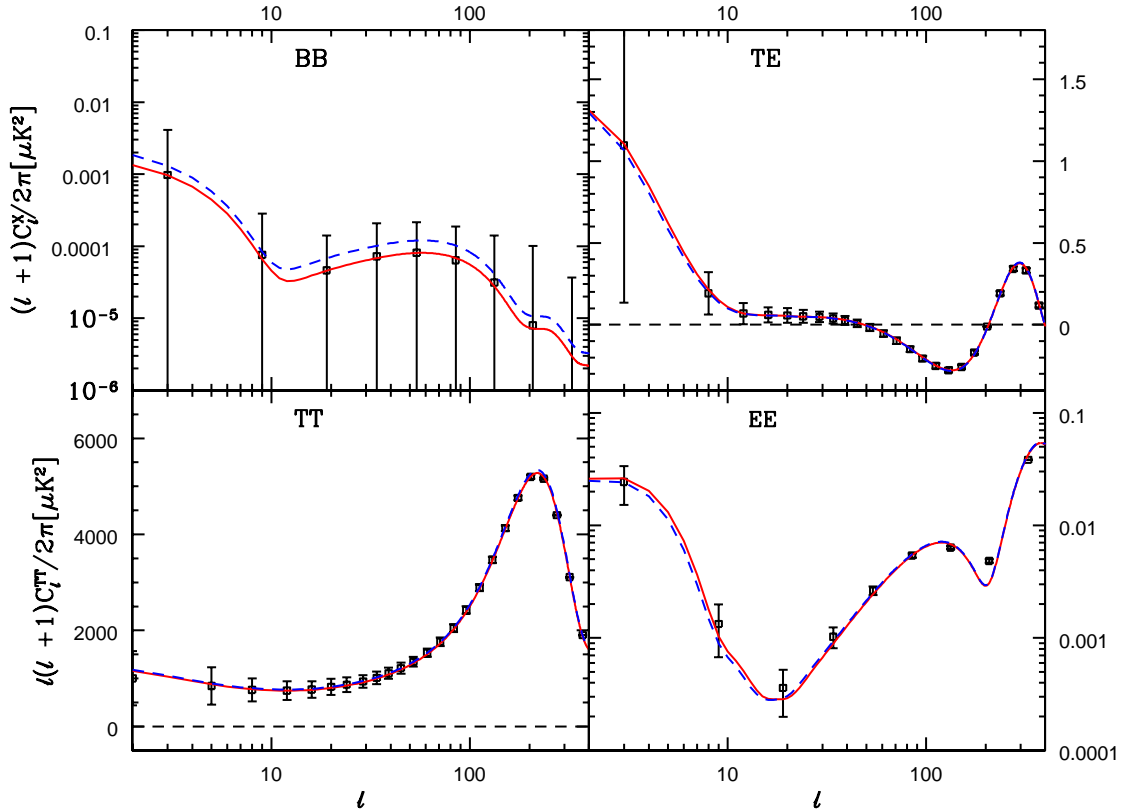}
\end{center}
\caption{\small Simulated recovery of the temperature and polarization angular power spectra from a 4-day first flight of \spider, with 100 and 145 GHz as in Table \ref{tbl:detectors}.  The simulations of time-ordered data  include 1/f noise with a 100 mHz knee frequency.  The half-wave plate is stepped each hour during the daytime scan mode and each night during nighttime spin mode. The sky signal is reconstructed from the time ordered data with a naively binned map-maker, power spectra are estimated with xfaster\cite{xfaster}.  The input model (solid curve) is the best fit $\Lambda$CDM + tensors of WMAP3.  The dashed curve is our own best fit spectrum with {\tt cosmomc} software \cite{2002PhRvD..66j3511L} - see Figure \ref{fig:params}.  These are expected to be conservative estimates of \spider\rq s sensitivity;  we expect to reduce the error in the lowest EE bin by $\sim$ 30\% with an optimal map maker.
}
\label{fig:performance}
\end{figure}

\begin{figure}[!t]
\begin{center}
\includegraphics[width=3in]{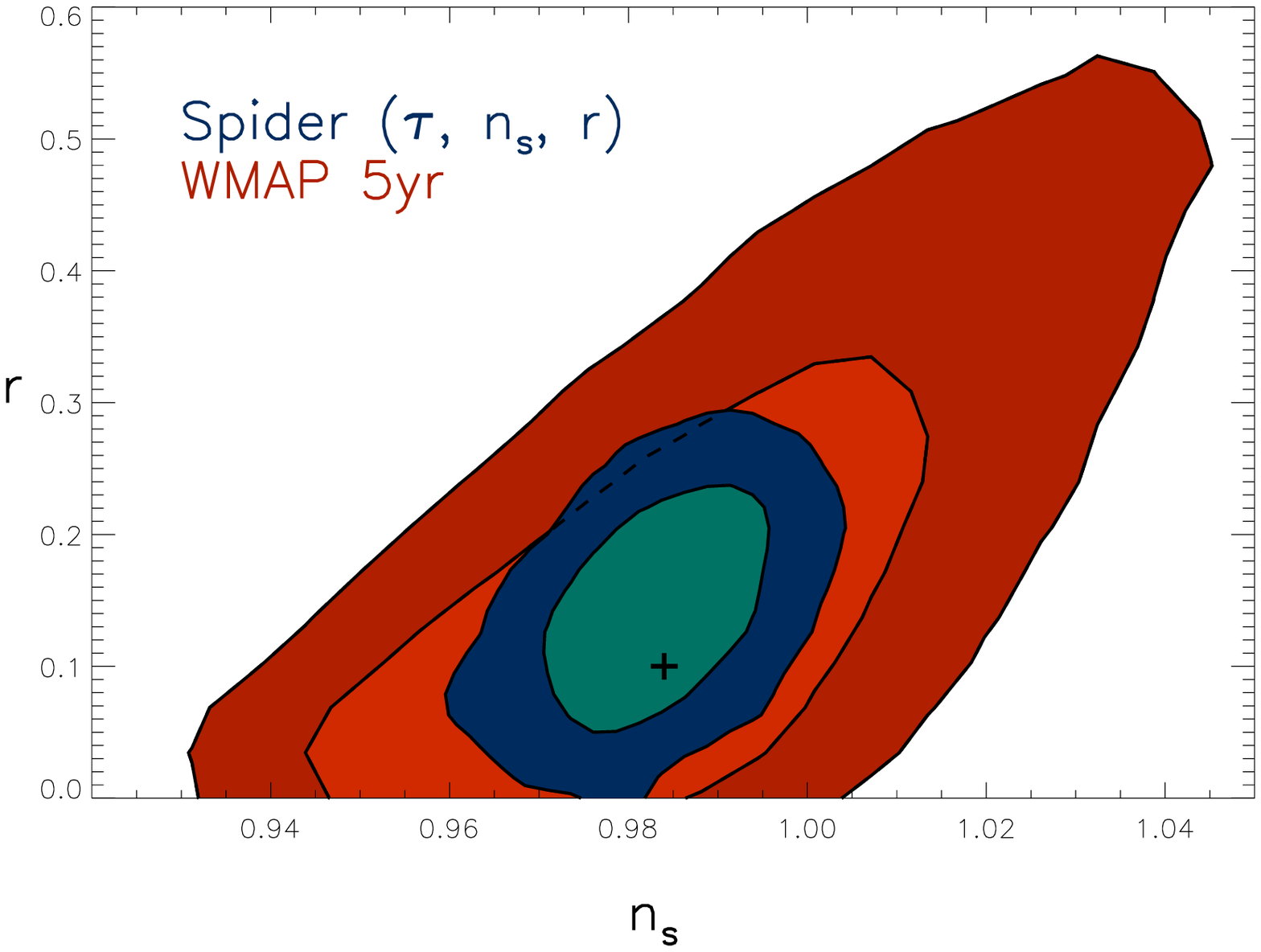}
\includegraphics[width=3in]{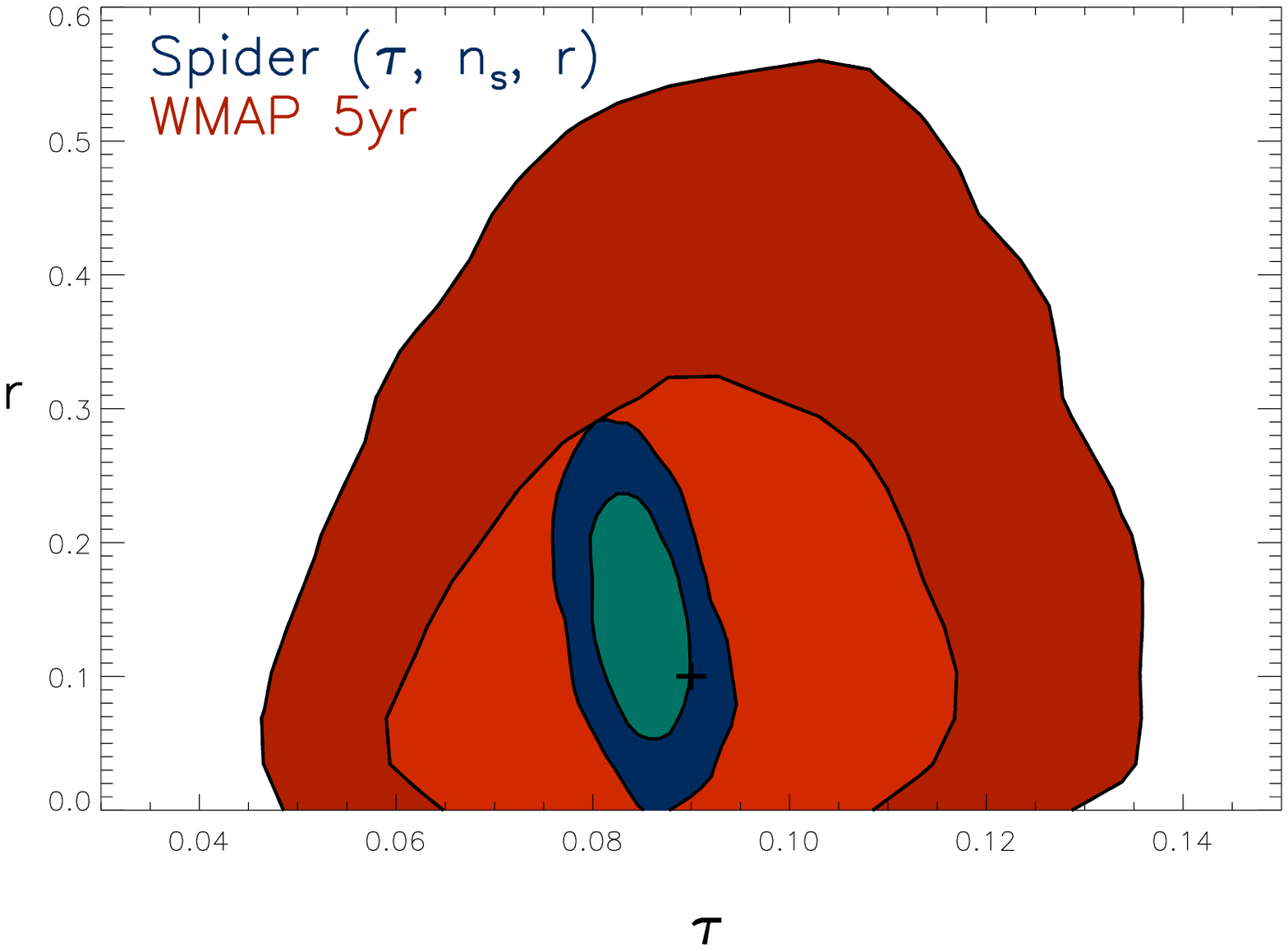}
\end{center}
\caption{\small Marginalized likelihood contours for WMAP 5-year and \spider\rq s first flight.  $r$ is the ratio of tensor perturbations to scalar perturbations, $\tau$ is the optical depth through reionization, and $n_s$ is the spectral index of scalar perturbations.  The $+$ represents the input value of the parameters; in this case $n_s=0.984$, $r=0.1$, $\tau=0.09$.  \spider\rq improvement comes from more sensitive polarimetery on large scales.
}
\label{fig:params}
\end{figure}

\subsection{The Epoch of Reionization}
\label{sec:reionization}

A variety of astrophysical measurements show that the Universe was ionized sometime before redshift of $z \sim 6$\cite{2004ApJ...617L...5M,2006PASJ...58..485T}, presumably due to ultraviolet photons from star formation.   Measurement of the large-scale polarization of the CMB provides a very sensitive measurement of the total optical depth of the reionized universe out to the surface of last scattering.   The five-year observations of WMAP give an optical depth of $\tau = 0.087 \pm 0.017$.  Assuming that  reionization occurred instantaneously, it occurred at a redshift $11.0 \pm 1.4$ \cite{2008arXiv0803.0586D}.   A deeper polarization map from \spider\rq s test flight will give a measurement of the optical depth that is many times more precise than WMAP\rq s 5-year measurement (Figure \ref{fig:params}).  The improvement comes from \spider\rq s deep measurement of polarization on the largest scales, a unique  measurement from a suborbital mission.

\subsection{Dust in the Galaxy}
\label{sec:dust}
The polarized signal from the galaxy at millimeter wavelengths is mainly from synchrotron at frequencies below 70 GHz and thermal emission by dust at higher frequency.  WMAP provides measurements of synchrotron emission\cite{2008arXiv0803.0715G} and in doing so, make a map of the large scale magnetic field of the galaxy\cite{2007ApJS..170..335P}.  Dust grains preferentially align along galactic magnetic fields, polarizing the thermal emission.   \spider\rq s frequency coverage is chosen to distinguish CMB polarization from dust\cite{2006SPIE.6267E..24M}, making \spider\ an ideal instrument for dust studies.  The degree of grain alignment (and hence polarization fraction in the \spider\ passbands) depends on the physical properties of the dust grains.    Additionally, using polarized dust emission to constrain the magnetic field strength on smaller scales in the galaxy gives insight into the role of magnetic fields in star formation. 

\subsection{Probing Inflation}
\label{sec:inflation}
A third science goal of \spider\ is to probe the energy scale of the inflationary period of the Universe by mapping the polarization of the CMB on very large scales.

Inflation is a postulated period of exponential expansion of the early universe, caused by an inflaton field with negative pressure.  Inflation solves problems in the standard model of cosmology, such as the horizon problem and the flatness problem, but the theory is most compelling as a means of seeding structure formation.  Quantum perturbations in the inflaton field expanded to cosmological sizes during inflation.  These inhomogeneities in the inflaton field drove density perturbations in matter and radiation, which in turn produced anisotropies in the CMB and eventually structure in the Universe.   The existence of acoustic peaks in the CMB anisotropy spectrum give circumstantial evidence to the inflationary scenario of structure formation\cite{2001PhRvD..63d2001L}.    

The imprint of gravitational waves on the CMB provides a direct way to probe inflation. Inflation produces perturbations in the metric tensor, effectively a background of super-horizon gravitational waves, that do not lead to structure formation, but do leave an imprint on the anisotropy and polarization of the CMB.   The gravitational wave background is the only cosmological means of generating B-mode fluctuations, and as such, is a unique signature of inflation.

The amplitude of the gravitational waves is parametrized with the tensor-to-scalar ratio $r$, the ratio of power in primordial tensor perturbations to power in scalar perturbations.  The amplitude of the gravitational waves is directly proportional to the energy scale of the inflaton particle before inflation.  In turn, the amplitude of the B-modes directly measures the energy scale of inflation.  The tightest limit on $r$ is from WMAP; assuming a power-law for the spectrum of tensor and scalar fluctuations, $r<0.43$\cite{2008arXiv0803.0586D}.  A long duration flight of \spider\ has excellent prospects of measuring the B-mode signal\cite{2006SPIE.6267E..24M}; however,  \spider\rq s first flight will not have long enough integration time to do so.

\section{INSTRUMENT}
\label{sec:instrument}

\begin{figure}[!t]
\begin{center}
\includegraphics[width=3in]{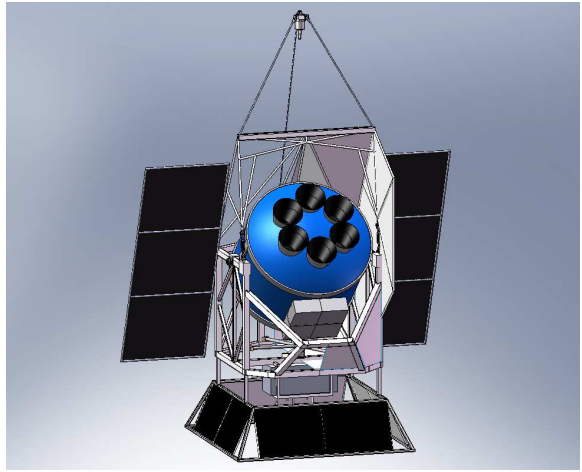}
\end{center}
\caption{\small
The \spider\ payload.  Six independent monochromatic telescopes are housed in a single long hold time cryostat.  Each telescope is fully baffled from radiation from the ground and balloon.  The gondola scans in azimuth with a reaction wheel and a motorized pivot.   The cryostat, mounted on bearings, can be adjusted in elevation.  Solar arrays provide power.
}
\label{fig:payload}
\end{figure}

The \spider\ payload is shown in Figure~\ref{fig:payload}.  The pointing and pointing reconstruction systems build on experience from the \boom\cite{2006A&A...458..687M} and \blast \cite{2007arXiv0711.3465P} designs.    The payload observes two ways: scanning in azimuth or spinning.  The inner frame is 
adjustable in elevation with a simple linear actuator similar to the \boom\ design.  Multiple tracking star cameras, rate gyros, differential GPS and a sun sensor provide pointing information.  The gondola is constructed from carbon fibre tubes to save mass.

\section{CRYOGENICS}
\label{sec:cryogenics}
The cryostat for the \spider\ instrument uses liquid helium-4 (LHe) to cool the instrument during its flight.  All six instrument inserts and the $\sim 1000$ litre LHe tank are contained in an outer vacuum vessel fabricated by Redstone Aerospace.  The primary LHe tank is maintained at 108 kPa and a small ($\sim$ 20 
litre) capillary-fed superfluid LHe tank will be controlled at a vapour 
pressure near 100 Pa. The inserts and the liquid cryogen tanks are surrounded by two concentric vapour-cooled shields and the inner tank is mechanically supported by G10 flextures. The use of staged vapour-cooled shields and radiation blockers reduces the radiative loading on the optics and detectors. Closed-cycle $^3$He sorption refrigerators, one per focal plane, will cool the detectors to 260 mK from the 1.5K base temperature.  The sorption fridges are cycled every 48 hours.

\section{OPTICAL DESIGN}
\label{sec:optics}
\subsection{Telescope}
The optical design is based on the successful \bicep\ telescope\cite{2006SPIE.6275E..51Y}.  Each telescope is a monochromatic, telecentric refractor with anti-reflection-coated polyethylene lenses, and is cooled to 4K. The aperture field distribution of the primary is smoothly tapered with an anodized 4K Lyot stop, reducing the detector background.

\subsection{Half-wave Plate}
\spider\ modulates the polarization of the incoming light with a stepped half-wave plate (HWP) at the telescope aperture.  Modulating the polarization mitigates systematic errors from asymmetric beams, instrumental polarization and relative gain uncertainty between detectors.  

A HWP placed at the aperture of the telescope rotates the angle of polarization sensitivity on the sky at four times the physical rotation rate of the HWP, while leaving the beams unchanged.   It also enables a full measurement of the sky polarization using each individual detector, eliminating or reducing many potential systematic effects.  

\begin{figure}[!t]
\begin{center}
\includegraphics[width=5in]{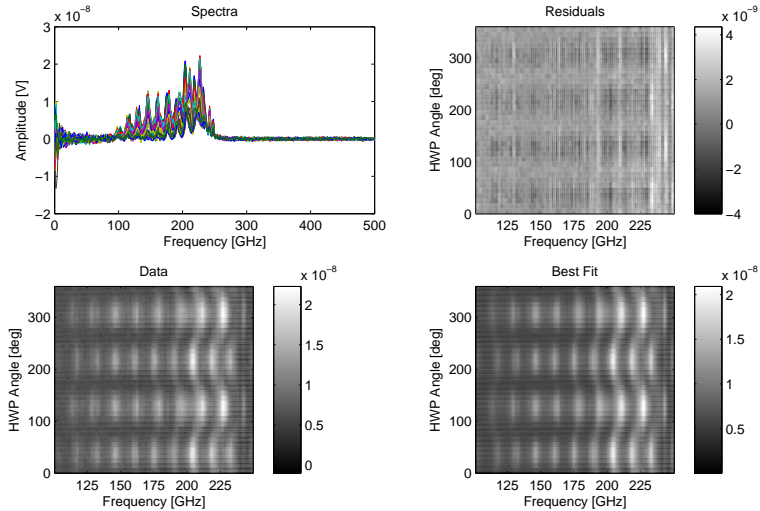}
\end{center}
\caption{\small
Measurements of the transmission of a polarized signal as a  
function of frequency and rotation angle for a single
birefringent sapphire plate appropriate for our 150GHz band, which  
will be used in \spider\rq s half-wave plate with an anti-reflection (AR) coating.
The data are well-described by a birefringence model that gives an index difference of $\Delta n =  
0.3350 \pm 0.0003$ between the fast and slow axes.
The model fit also recovers the angle of the crystal axes to 0.1$^\circ$  
precision.  The addition of an AR-coat greatly reduces the
channel spectra evident in band.}
\label{fig:hwp}
\end{figure}

\spider\rq s single-frequency telescopes simplify the HWP design and
implementation. A single birefringent sapphire wave plate coated with
a single layer of fused quartz on each side gives very good (band average of
96.8\% modulation efficiency) performance over a 25\% bandwidth.  Measurements of a prototype HWP with no anti-reflection coating (Figure~\ref{fig:hwp}) show excellent polarization modulation across the \spider\ band.

\section{DETECTORS AND READOUT}
\label{sec:detectors}

\subsection{Detectors and SQUIDs}
\begin{figure}[!t]
\begin{center}
\includegraphics[width=0.55\textwidth]{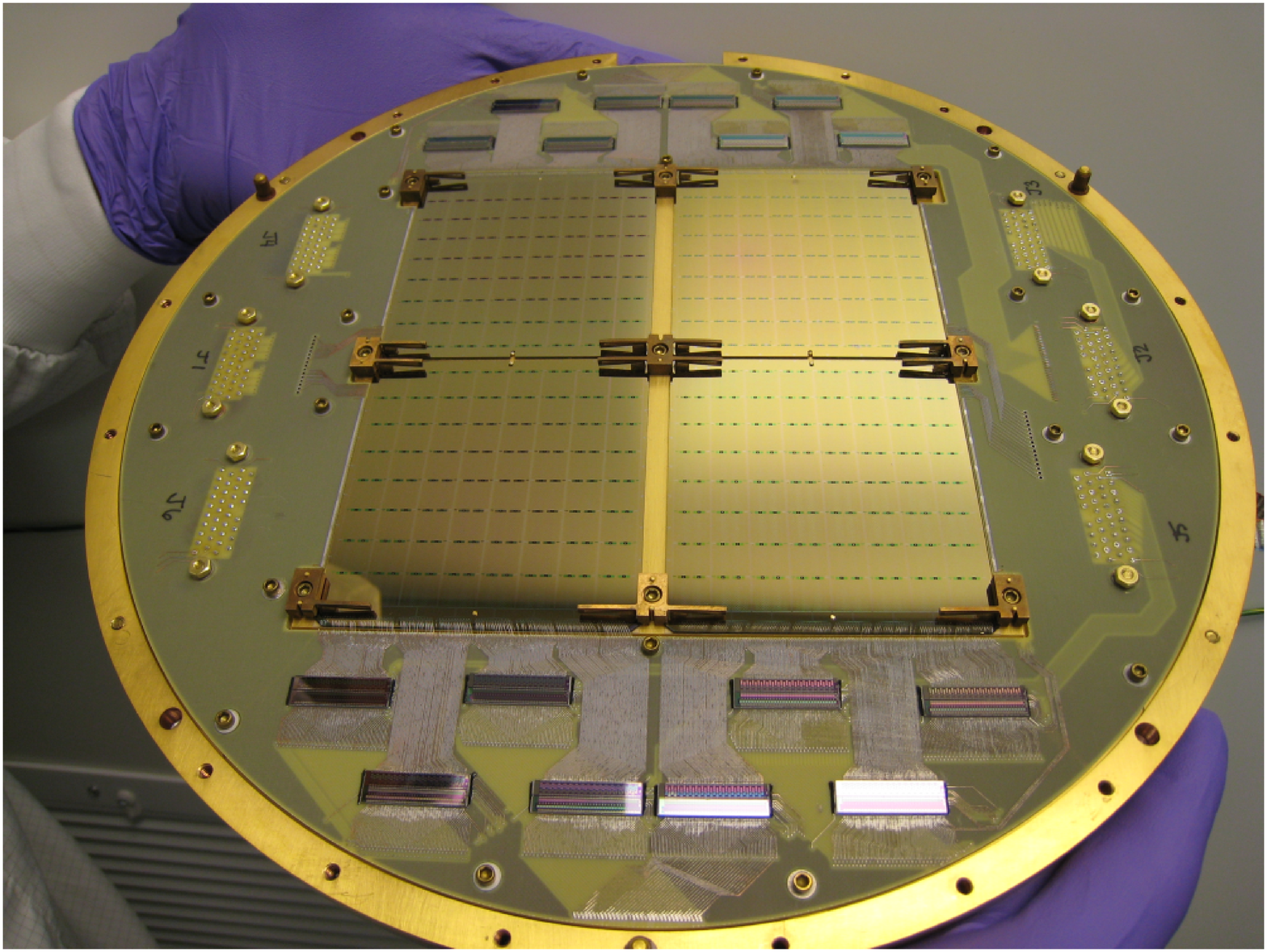}
\hfill
\includegraphics[width=0.4\textwidth]{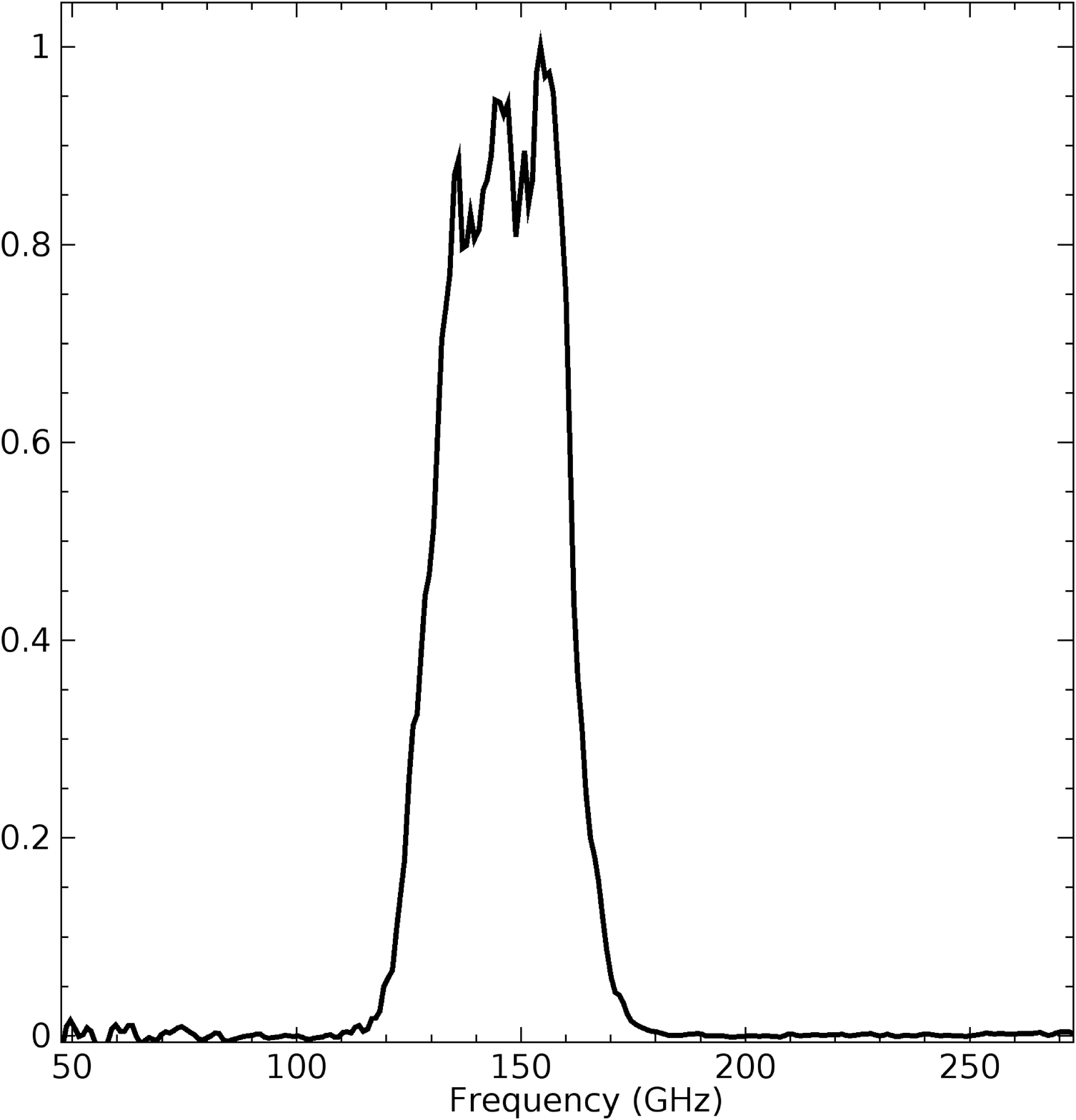}
\end{center}
\caption{\small
Left: A prototype \spider\ focal plane unit, consisting of four detector wafers.  Each wafer comprises 64 spatial pixels, sensing two polarizations each, and read out by 128 detectors.  The smaller chips on the periphery are NIST column multiplexer chips, each reading out 32 detectors.  Light enters through a square aperture under the tiles, back illuminating the detectors.  A niobium plate covers the entire assembly and serves as a superconducting magnetic shield as well as a backshort surface for the antennae.  Right: the passband of a 145 GHz device.  This passband is of the device alone, the \spider\ optical train is not yet mated to the detector.
}
\label{fig:detectors}
\end{figure}

\spider\ uses a large-format array of antenna-coupled bolometers\cite{goldin:251}, shown in
Figure \ref{fig:detectors}. Each spatial pixel consists of a $16\times 16$ phased array of slot dipole
antennae.  The pixel\rq s radiation pattern is defined by the
coherent interference of the antenna elements, yielding a
$\sim 13^\circ$~{\sc FWHM} beam with a first sidelobe
at -15 dB when the 256 elements are fed with equal weights.  Each spatial pixel has two orthogonally polarized antennae.\footnote{Each detector will be able to measure Q and U independently, but this is due to the use of a HWP and scan strategy to modulate the angle of polarization sensitivity.  The dual polarization design of the pixels is primarily useful for high device packing density.}
With the radiation pattern of the antenna array terminated on the 4K  stop, the system produces highly symmetrical beams on the sky with low 
cross-polarization. The optical power is transmitted via superconducting
microstrip\cite{vayonakis:539} to superconducting transition-edge sensors
(TESs)\cite{2003SPIE.4855..318H}.

The array design has a bandwidth of close to 30\%, set by the slot
length.  Optical band definition is provided by a combination of in-line
microstrip filters and cryogenic metal-mesh optical filters, the combination of which provide sharp band edges and no intrinsic out-of-band response (see Figure~\ref{fig:detectors}).

The TESs have strong electro-thermal feedback and low heat capacity, providing an extremely linear 
and fast bolometer~\cite{1995ApPhL..66.1998I, 1996ApPhL..69.1801L,1999ApPhL..74..868G}. Thermal
isolation of the TES films is provided by \sini support legs.  Optical
power is deposited in the TES by terminating the superconducting
microstrip with a long gold meander on the same thermal island as the
TES.

The antenna-coupled design is entirely photolithographically
fabricated, greatly simplifying production and the uniformity of large-format arrays.  The integrated components are mechanically robust, and immune to differential thermal contraction.
The densely populated antennae allow a very efficient use of the focal plane
area.  The antenna-coupled architecture has been successfully 
demonstrated~\cite{2003SPIE.4855..318H, 2004PhDT........24H}. We have thoroughly characterized a series of
end-to-end prototype detectors at 96 and 150 GHz.  Radiation patterns, measured with both optically modulated thermal sources and monochromatic sources, confirm the theoretical predictions.  The inline filters show 
well defined bands with $\sim$30\% bandwidth.

The distribution of the pixels and the per-pixel sensitivities are
indicated in Table~\ref{tbl:detectors}. In the CMB science bands (100 and 150~GHz), sensitivities are dominated by photon and phonon noises which contribute roughly equally.  Johnson and amplifier
noises are negligible.

\begin{table*}[t]
\begin{center}
\begin{tabular}{|c|c|c|c|c|c|c|}
\hline
Band  &  & Beam & Number of &    & Single-Detector & Instrument \\
Centre  & Bandwidth & FWHM & Spatial & Number of & Sensitivity & Sensitivity \\
(GHz) & (GHz) & (arcmin) & Pixels & Detectors & ($\mukcmbrts$)& ($\mukcmbrts$) 
    \\ \hline\hline
100 & 24 & 58  & ($2\times$)144 & ($2\times$)288 & 100 &  4.2 \\
145 & 35 & 40  & ($2\times$) 256 & ($2\times$) 512 & 100 &  3.1 \\ 
225 & 54 & 26  & 256 & 512 & 204 &  9.0 \\
275 & 66 & 21  & 256 & 512 & 351 &  15.5 \\ \hline\hline
\end{tabular}
\end{center}
\caption{\small Observing bands, pixel and detector counts, and single-detector
and instrument sensitivities. The latter is obtained by dividing the
single-detector sensitivity by $\sqrt{N_\textrm{ det}}$. A total of 2624
detectors are distributed between the six telescopes, with two telescopes in each of the 100 GHz and 145 GHz bands.}
\label{tbl:detectors}
\end{table*}

 The TES sensors will be read out using superconducting quantum
interference device (SQUID) current amplifiers with time-domain
multiplexing.  Ambient temperature multi-channel electronics (MCE) \cite{2008JLTP..151..908B} which work in concert with time-domain multiplexers~\cite{1999ApPhL..74.4043C, 2003RScI...74.3807D,
2003RScI...74.4500R, 2004NIMPA.520..544I}.   The MCE were initially developed for SCUBA2\cite{2006SPIE.6275E..45H} and are used as read-out electronics on many CMB and sub-mm
astronomy receivers (e.g. ACT \cite{2006AAS...20924003F}, C$_\ell$OVER \cite{2006SPIE.6275E..69A}, BICEP II and SPUD\cite{2006AAS...209.1105K}).

\subsection{Magnetic Shielding}
Magnetic shielding is a critical requirement for \spider\rq s receiver.   We use a combination of high-permeability and superconducting shields to achieve the necessary reduction in magnetic field.  A SQUID amplifier measures current via the magnetic flux in its input inductor coil, therefore it also responds to variations in the ambient magnetic field (albeit reduced by a quadrupole gradiometric input coil).  Additionally, the critical temperature ($T_c$) of the TES devices depend on the applied field.  These susceptibilities can be measured in the lab, allowing us to set a specification for the required level of magnetic shielding.  Our requirement on spurious field signal is that it is less than the expected map rms on $1^\circ$ pixels.

We distinguish between two types of sources of magnetic fields: 1) "ambient" fields over which we have no control; and 2) fields generated locally by motors and electronics on the instrument or in the nearby environment. A typical value of Earth's magnetic field 50 $\mu T$.  The local magnetic field due to \spider\rq s azimuth drive will be less than 1 $nT$ at the location of the focal plane.  We therefore focus on shielding of the ambient field.

We have measured a 1 mK/50 $\mu T$ field dependence of $T_c$ for our Ti TES devices.  Table \ref{tbl:B_sensitivity} presents the ambient fields, field sensitivity, and resulting shielding requirements for \spider. In the case of the SQUID susceptibility, the parameter in question is "effective area", which converts from applied field to magnetic flux.  The current NIST design has an effective area of $(12 \mu m)^2$.  We show in Table \ref{tbl:B_sensitivity} how this susceptibility feeds through the system.  If insufficient, we also try, at some cost in complexity: a) using driven coils to null Earth's field or b) use AC modulation of the TES bias to move the science signal outside the field pickup band.
\begin{figure}[!t]
\begin{center}
\includegraphics[width=4in]{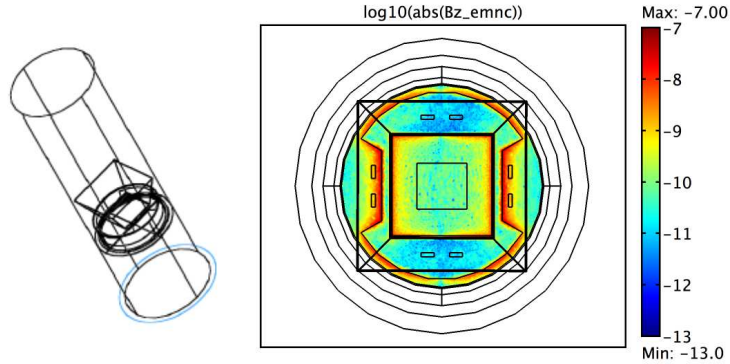}
\end{center}
\caption{ \small {\em Left:} Magnetic shielding design for \spider. The \spider\ design incorporates a single layer of cryoperm at 4K (long tube, 1.25m long), a sub-K superconducting box with a flared opening through which light enters, and a superconducting plane just under the array. {\em Right:} the z-component of the magnetic field in the focal plane, 500 $\mu$m above the superconducting plane. A 50 $\mu$T field has been applied at a $45^\circ$ angle to the optical axis.  The hourglass shape is the outline of the superconducting plane -- the vertical field is highest at the edges where it goes around the superconducting plane.  The large square in the centre represents the TES focal plane.  The rectangles at the top and bottom of the plot are indicative of where the SQUIDs will be placed. We find a residual field of $<$ 300 pT over most of the TES focal plane and at the few pT level at the SQUIDs.  }
\label{fig:magshielding}
\end{figure}

We have designed a magnetic shielding using COMSOL finite element calculations\cite{Chui:MagneticShielding} (see Figure \ref{fig:magshielding}); the residual field levels meet the TES and SQUID susceptibility goals listed in Table \ref{tbl:B_sensitivity}.   Additional rejection of field signals based on the use of dark pixels or common-mode signals among many science pixels will provide a large safety margin over these estimates.

\begin{table}

\begin{tabular}{|p{1.9in}|l|p{0.3in}|p{1.9in}|l|}
\cline{1-2}
\cline{4-5}
Desired Depth & 0.5 $\mu K_{CMB}$ & & Desired Depth & 0.5 $\mu K_{CMB}$ \\
\cline{1-2}
\cline{4-5}
Ambient Field & 50 $\mu T$ & & Ambient Field & 50 $\mu T$ \\
\cline{1-2}
\cline{4-5}
TES G & 20 $pW/K$ & & SQUID field/ TES current & 0.5 $pA/pT$ \\
\cline{1-2}
\cline{4-5}
TES $T_{bolo}$/$T_{CMB}$ responsivity & 7 $mK / K_{CMB}$ & & TES current/$T_{CMB}$ responsivity & 0.7 $pA / \mu K_{CMB}$ \\
\cline{1-2}
\cline{4-5}
TES field sensitivity & 2.7 $\mu K_{CMB} / nT$ & & Field sensitivity via SQUID & 0.7 $\mu K_{CMB} / pT$ \\
\cline{1-2}
\cline{4-5}
Residual field requirement&  180 $pT$ & & Residual field requirement &  0.7 $pT$ \\
\cline{1-2}
\cline{4-5}
Attenuation requirement & $3 \times 10^5$ & & Attenuation requirement & $7 \times 10^7$ \\
\cline{1-2}
\cline{4-5}
\end{tabular}

\caption{\small {\em Left:} Summary of TES magnetic field sensitivity and shielding requirements.  We consider the 150 GHz band with optical efficiency 0.4 and a TES field sensitivity of 1 mK/50 $\mu$T. {\em Right:} Summary of SQUID magnetic field sensitivity.  The input coil / SQUID mutual inductance is 275 pH.  
}
\label{tbl:B_sensitivity}
\end{table}

\section{FOREGROUNDS AND SYSTEMATICS}
\label{sec:foregrounds_systematics}

Polarized emission from the galaxy confuses measurements of CMB polarization, and is likely to set the ultimate limit on measurement of B-modes.  The bandpasses for \spider\ are selected to distinguish E- and
B-mode signals from galactic emissions.  The band selection strategy has not changed since Montroy et al (2006) and we refer the reader to that work\cite{2006SPIE.6267E..24M}.

Instrumental systematic effects present a huge challenge for the measurement of nanoKelvin polarization signals on the sky \cite{2003PhRvD..67d3004H,2007A&A...464..405R}.   MacTavish et al. (2007) \cite{2007arXiv0710.0375M} investigate systematics in the specific case of \spider\ using a full data analysis pipeline.  This set of simulations set specifications on instrumental design and on in-flight or pre-flight knowledge of various instrumental parameters (Table \ref{tbl:specs}).

\begin{table*}
\centering
\begin{tabular}{|c|c|c|}
\hline
\small
\space
{\bf Systematic} & {\bf Experimental Spec.} &
{\bf Comments} \\
\hline
Receiver 1/f knee & $< 200$ mHz &  for 110 degrees per second gondola spin\\
\hline
Receiver 1/f knee & $< 100$ mHz &  for 36 degrees per second gondola spin\\
\hline
Pointing Jitter  & $< 10'$ & sufficient for $\ell < 50$\\
\hline
Absolute Pol. Angle Offset & $< 0.25^\circ$  &  \\
\hline
Relative Pol. Angle Offset & $< 1^\circ$ &  \\
\hline
Beam Centroid Positions & $< 1'$ &  sufficient for $\ell < 30$ \\
\hline
Optical Ghosting & $< 2\%$ &  \% TOD contamination\\
\hline
Calibration Drift & $< 3.0\%$ & in phase  \\
\hline
Calibration Drift & $< 0.1\%$ & out of phase \normalsize \\
\hline
\end{tabular}
\caption{\small Summary of experimental specifications based on simulation results, a reproduction of Table 2 of MacTavish et al (2007)..    Realistic-amplitude, time-varying systematics are 
simulated in time streams.  Maps are reconstructed without any attempt to
correct for the systematic errors.  Experimental specifications are set by limiting the
allowed systematic residual level to a factor of $\sim10$ smaller than the 
B-mode signal for $r = 0.01$.  The nominal operating mode is a 36 dps gondola spin rate, with
the half-wave plate stepping $22.5^{\circ}$ once per day, with 10
iterations of a Jacobi iterative map-maker\cite{2007A&A...470..771J}.  Each of these specifications is easily met by \spider.
}\label{tbl:specs}
\end{table*}

\section{CONCLUSIONS}
\label{sec:conclusions}

\spider\ is unique as a sub-orbital millimeter-wave polarimeter that can observe a large fraction of the sky.  
 Large-format multiplexed TES bolometer arrays provide excellent raw sensitivity,  and the refracting optical design and a rotating half-wave plate will minimize systematics.  In its first flight, \spider\ will make a precise measurement of the optical depth of reionization.  Later flights will measure the B-mode signal from primordial gravitational waves. 

\spider\ acts a pathfinder to the proposed CMBpol space mission more than any other planned B-mode experiment.   \spider\ will test a receiver with a large format array and multiplexed readout electronics in a high-altitude environment as well as observing strategies and systematics of deep polarimetery on a large fraction of the sky.  The design of \spider emphases lightweighting, power savings, and autonomous operation all of which are directly applicable to a space mission.

\acknowledgments 
We acknowledge funding from NASA (grant number NNX07AL64G), the Gordon and Betty Moore Foundation, and NSERC.  We are grateful to Danny Ball and the Columbia Scientific Ballooning Facility for assistance with flight planning.

%%%%% References %%%%%
\bibliography{Spider_2008}
\bibliographystyle{spiebib}

\end{document}